\documentclass[modern]{aastex631}

\defcitealias{paper2}{W23}

\submitjournal{RNAAS}

\begin{document}

\title{The JWST Resolved Stellar Populations Early Release Science Program III: Photometric Star-Galaxy Separations for NIRCam}

\author[0000-0003-1634-4644]{Jack T. Warfield}
\affiliation{Department of Astronomy,
The University of Virginia,
530 McCormick Road,
Charlottesville, VA 22904, USA
}
\author[0000-0002-3188-2718]{Hannah Richstein}
\affiliation{Department of Astronomy,
The University of Virginia,
530 McCormick Road,
Charlottesville, VA 22904, USA
}
\author[0000-0002-3204-1742]{Nitya Kallivayalil}
\affiliation{Department of Astronomy,
The University of Virginia,
530 McCormick Road,
Charlottesville, VA 22904, USA
}
\author[0000-0002-2970-7435]{Roger E. Cohen}
\affiliation{Department of Physics and Astronomy,
Rutgers University,
136 Frelinghuysen Road, Piscataway, NJ 08854, USA
}
\author[0000-0002-1445-4877]{Alessandro Savino}
\affiliation{Department of Astronomy,
University of California, Berkeley,
Berkeley, CA 94720, USA
}

\author[0000-0003-4850-9589]{Martha L. Boyer}
\affiliation{Space Telescope Science Institute, 3700 San Martin Drive, Baltimore, MD 21218, USA}

\author[0000-0001-9061-1697]{Christopher T. Garling}\affiliation{Department of Astronomy,
The University of Virginia,
530 McCormick Road,
Charlottesville, VA 22904, USA
}

\author[0000-0002-5581-2896]{Mario Gennaro}
\affiliation{Space Telescope Science Institute, 3700 San Martin Drive, Baltimore, MD 21218, USA}
\affiliation{The William H. Miller {\sc III} Department of Physics \& Astronomy, Bloomberg Center for Physics and Astronomy, Johns Hopkins University, 3400 N. Charles Street, Baltimore, MD 21218, USA}

\author[0000-0001-5538-2614]{Kristen B. W. McQuinn}
\affiliation{Department of Physics and Astronomy,
Rutgers University,
136 Frelinghuysen Road, Piscataway, NJ 08854, USA
}

\author[0000-0002-8092-2077]{Max J. B. Newman}
\affiliation{Department of Physics and Astronomy,
Rutgers University,
136 Frelinghuysen Road, Piscataway, NJ 08854, USA
}

\author[0000-0003-2861-3995]{Jay Anderson}
\affiliation{Space Telescope Science Institute, 3700 San Martin Drive, Baltimore, MD 21218, USA}

\author[0000-0003-0303-3855]{Andrew A. Cole}
\affiliation{School of Natural Sciences, University of Tasmania, Private Bag 37, Hobart, Tasmania 7001, Australia}

\author[0000-0001-6464-3257]{Matteo Correnti}
\affiliation{Space Telescope Science Institute, 3700 San Martin Drive, Baltimore, MD 21218, USA}

\author[0000-0001-8416-4093]{Andrew E. Dolphin}
\affiliation{Raytheon Technologies, 1151 E. Hermans Road, Tucson, AZ 85756, USA}
\affiliation{Steward Observatory, University of Arizona, 933 N. Cherry Avenue, Tucson, AZ 85719, USA}

\author[0000-0002-7007-9725]{Marla C. Geha}
\affiliation{Department of Astronomy, Yale University, New Haven, CT 06520, USA}

\author[0000-0002-4378-8534]{Karin M. Sandstrom}
\affiliation{Center for Astrophysics and Space Sciences, Department of Physics, University of California San Diego, 9500 Gilman Drive, La Jolla, CA 92093, USA}

\author[0000-0002-6442-6030]{Daniel R. Weisz}
\affiliation{Department of Astronomy,
University of California, Berkeley,
Berkeley, CA 94720, USA
}

\author[0000-0002-7502-0597]{Benjamin F. Williams}
\affiliation{Department of Astronomy, University of Washington, Box 351580, U.W., Seattle, WA 98195-1580, USA}

\begin{abstract}
We present criteria for separately classifying stars and unresolved background galaxies in photometric catalogs generated with the point spread function (PSF) fitting photometry software \texttt{DOLPHOT} from images taken of Draco II, WLM, and M92 with the Near Infrared Camera (NIRCam) on JWST. Photometric quality metrics from \texttt{DOLPHOT} in one or two filters can recover a pure sample of stars. Conversely, colors formed between short-wavelength (SW) and long-wavelength (LW) filters can be used to effectively identify pure samples of galaxies. Our results highlight that the existing \texttt{DOLPHOT} output parameters can be used to reliably classify stars in our NIRCam data without the need to resort to external tools or more complex heuristics. 
\end{abstract}

%% https://astrothesaurus.org

\section{Introduction} \label{sec:intro}

The JWST Resolved Stellar Populations ERS program (PID 1334; \citealt{paper2}, hereafter \citetalias{paper2}) was approved to develop data reduction methods and tools to investigate strategies for future observations of resolved stars in nearby galaxies and star clusters. In this note, we present photometric cuts in color and in the \texttt{DOLPHOT} \citep{dolphin2000b, dolphin2016} parameters \texttt{ROUND}, \texttt{SHARP}, \texttt{CROWD}, \texttt{FLAG}, and \texttt{OBJECT\_TYPE} that can be used to recover pure samples of stars or background galaxies from our NIRCam observations of Draco II, WLM, and M92. 
Separating out likely stars from the background
%(or, inversely, removing sources that are confidently galaxies)
is necessary for accurately analyzing a target's true stellar population---particularly important for cases such as modeling a color-magnitude diagram (CMD)
% ; e.g., for the initial mass function or star formation histories)
or measuring proper motions.
%in the case of ultra-faint dwarfs, such as Draco II, where there are so few stars.
%Alternatively, a pure catalog of background galaxies may be useful for building a background reference frame for proper motion measurements.

The \texttt{DOLPHOT} parameters presented herein can be used to recover star-like objects from the photometric catalog produced for each target. Additionally, for fields without significant crowding, colors formed through the combination of SW and LW filters can be used to separate galaxies from stars in color-color and CMD space, especially at high signal-to-noise ratios (SNR).
These proof-of-concept criteria demonstrate how long-baseline colors and photometric morphology parameters can be employed to accurately determine object types, and are offered as starting points for refining photometric catalogs from any JWST NIRCam observations.

\section{Observations and Photometry} \label{sec:obs}

The design of the ERS observational program is given in \citetalias{paper2}. For the SW channel, all three targets have images in the F090W and F150W filters. 
% Draco II
% the NIRCam observations placed the center of Draco II in the gap between modules A and B, allowing a large areal overlap with archival Hubble Space Telescope (HST) imaging (HST-GO-14734; PI N. Kallivayalil) from March 2017.
The LW filters for Draco II were F360M and F480M. The integration time was 11,810 seconds per filter for F090W and F277W and 5883 seconds for F150W and F444W.
% WLM
For WLM, the LW filters were F250M and F430M.
The integration times were 30,492 seconds in F090W and F430M and 23,706 seconds in F150W and F250M.
% The NIRCam observations for WLM were oriented such that one module would overlap both multi-waveband HST and ALMA archival observations and the other would overlap with deep HST/ACS imaging (HST-GO-13768; PI D. Weisz). 
% The SW filters were again F090W and F150W, and the LW filters were F250M and F430M. The integration times were 30,492 seconds in F090W and F430M and 23,706 seconds in F150W and F250M.
% M92
For M92, LW observations were taken in F227W and F444W. The integration time chosen for all 4 filters was 1245 seconds.
% The observations for M92 centered the globular cluster between modules A and B, thus avoiding the saturated and highly-crowded center. The SW filters were F090W and F150W, while F227W and F444W were chosen for the LW channel. The integration time chosen for each filter was 1245 seconds.

Observations were reduced with the new NIRCam module of \texttt{DOLPHOT}\footnote{\href{http://americano.dolphinsim.com/dolphot}{http://americano.dolphinsim.com/dolphot}} \citep{dolphin2000b, dolphin2016}, whose specific functionality is discussed further in \citetalias{paper2}. 
\texttt{DOLPHOT} is a PSF photometry package capable of identifying point sources and measuring their positions and magnitudes from individual exposures. 
This module includes a preprocessing routine to apply the appropriate masks and pixel area maps provided by the Space Telescope Science Institute (STScI) pipeline for NIRCam, along with functions for using image metadata and camera-specific PSF models.
%As \texttt{DOLPHOT} photometry can be heavily affected by parameter choices,
Our initial reduction parameter choices were based on those used by the Panchromatic Hubble Andromeda Treasury for target imaging \citep{Williams2014}.
%For this note, we only consider objects in each field for which a magnitude $<50$ has been recorded in all available filters, and for which the SNR in all filters is $>20$. 
For our stellar classification, we only consider sources for which ${\rm SNR} > 10$ in the F090W and F150W filters. For our galaxy classification, we only consider sources for which ${\rm SNR} > 10$ in all filters.

\section{Cuts for Stars} \label{sec:stars}

\begin{figure*}
    \plotone{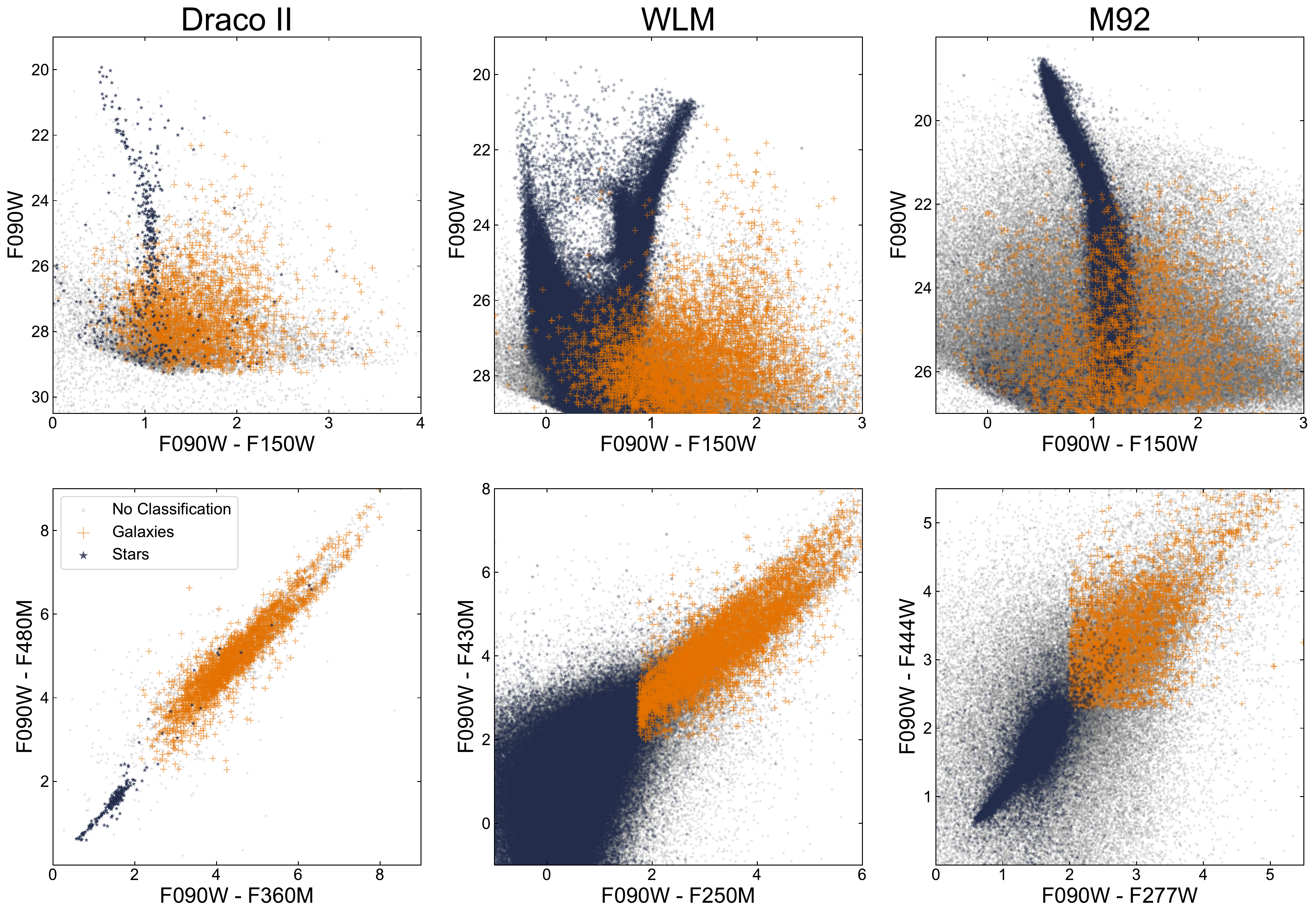}
    \caption{Blue stars represent objects classified as stars and orange pluses as galaxies. Gray circles are objects that fit our general SNR cuts, but fail either classification. The top row of plots shows CMDs for each field in the sample. The bottom row shows examples of color-color plots for each field. \label{fig}}
\end{figure*}
We find that a general cut for a purity-driven sample of stars can be made with:
% Jack: Stringent cut from Roger---we could also (or instead) include the looser cut, but I think that it is probably better to provide cuts for more "pure" samples?
% Jack: Need to convert the object type to the chi cuts (and the flags?)
\begin{verbatim}
    F090W_SHARP^2 & F150W_SHARP^2 <= 0.01,
    F090W_CROWD & F150W_CROWD <= 0.5,
    F090W_FLAG & F150W_FLAG <= 2, and
    OBJECT_TYPE <= 2.
\end{verbatim}
Here, we only consider the F090W and F150W filters. 
% This is both due to the performance of these filters individually for probing stellar populations as well as the usefulness of this color baseline.
This choice is favorable due to the sensitivities of the individual filters for probing stellar populations, as well as the usefulness of the color baseline for experiments such as deriving star formation histories. 
However, in principle, this set of cuts can be done with single-filter photometry.
% showing parameters coming out of DOLPHOT are doing what they're supposed to do (SHARP, CROWD), to relatively high precision (Chris will write)

%As an audit of the reliability of this classification, we have visually checked the profiles of a random sample of stars in each field that passes this set of cuts. Through this lens, the status of objects fainter than ${\rm F150W} \approx xx.x$ as stars is precarious, with a dwindling distinction between unresolved galaxies and stars below this magnitude. Nonetheless, this cut is able to successfully reveal the evolutionary sequence of the high-SNR stars in each target galaxy/cluster.

\section{Cuts for Galaxies} \label{sec:gals}

Even in the NIR, distant galaxies tend to appear redder than stellar point sources.
While we see in Figure \ref{fig} (top) that the stellar sequence is entwined with the background galaxies, it is possible to disentangle these objects from each other by constructing long-baseline colors.
%Using the CMDs for each target,
Replacing the $x$-axis color with both possible ${\rm F090W - LW}$ color combinations in each target's CMD, and drawing cuts for objects redder than the edge of each field's stellar sequence, we define cuts of $({\rm F090W-F360M} > 2.1)$ and $({\rm F090W-F480M} > 2.1)$ for Draco II, $({\rm F090W-F250M} > 1.75)$ and $({\rm F090W-F430M} > 2)$ for WLM, and $({\rm F090W-F277W} > 2)$ and $({\rm F090W-F444W} > 2.3)$ for M92. Additionally, we defined cuts on \texttt{SHARP} and \texttt{ROUND} to select extended sources:
\begin{verbatim}
    F150W_SHARP^2 > 0.05 & F090W_SHARP^2 > 0.05,
    |F090W_SHARP^2 - F150W_SHARP^2| < 0.15,
    F090W_ROUND > 0.05 & F150W_ROUND > 0.03, and
    |F090W_ROUND - F150W_ROUND| < 0.2.
\end{verbatim}
These morphological cuts are effective at further purifying our selection; 
of objects that satisfy the color cuts for Draco II, WLM, and M92, these morphological cuts remove about 50\%, 25\%, and 10\% of those objects, respectively.

The color-color panels in Figure \ref{fig} (bottom) show the varying separation between each field's probable stars and galaxies. This separation is especially apparent for Draco II, where we also see the selection effect of only detecting the brightest stars at this SNR. In the case of WLM, we see greater overlap between the two classifications in this space. For both Draco II and WLM, we have visually verified that most of these sources appear likely galaxies based on their extended morphologies in the images and found that many of the spurious classifications correspond with diffraction artifacts from saturated stellar sources.
%, not with stellar point-sources.

%For M92, this separation is less obvious.
%Though the ${\rm F090W-F277W}$ color should, theoretically, provide a better baseline than ${\rm F090W-F250M}$,
For M92, the extreme crowding in this field leads to a much smaller sample of detected background galaxies. Furthermore,
it is difficult to confirm these classifications from the images
%crowding makes it difficult to confirm these classifications from the images,
%since sources tagged as galaxies are visually obscured by foreground artifacts.
since the foreground visually obscures sources tagged as galaxies.

\section{Conclusions}

From our \texttt{DOLPHOT} analysis of NIRCam observations of Draco II, WLM, and M92, we conclude that:
\begin{enumerate}
    \item \texttt{DOLPHOT} output parameters, even from single-filter photometry, are effective at cutting a catalog down to a pure sample of stars.
    \item Colors made by combining SW and LW filters can separate background galaxies from stars. These baselines should be maximized for experiments where this separation is desired.
    \item \texttt{DOLPHOT} output parameters can alone be used to distinguish background galaxies, and greatly augment the sample purity when used in tandem with color cuts.
    %\item Morphological cuts made with \texttt{DOLPHOT} parameters can be effective alone for identifying a catalog of background galaxies or used in tandem with long-baseline color cuts to greatly augment the catalog's purity; and,
    \item For crowded fields, diffraction artifacts from bright stars may significantly affect one's ability to recover background galaxies from NIRCam catalogs.
\end{enumerate}

\begin{acknowledgements}
This work uses observations made with the NASA/ESA/CSA JWST. The data were obtained from MAST at STScI, operated by the Association of Universities for Research in Astronomy, Inc. (AURA), under NASA contract NAS 5-03127 for JWST. These observations are associated with program DD-ERS-1334. This program also benefits from recent \texttt{DOLPHOT} development work using observations made with the NASA/ESA HST obtained from STScI (HST-GO-15902), and operated by AURA under NASA contract NAS 5–26555.  
\end{acknowledgements}

\textit{Facilities:} JWST (NIRCam), MAST

\textit{Data:} NIRCam observations \dataset[10.17909/qnej-tt71]{http://dx.doi.org/10.17909/qnej-tt71}

\textit{Software:} 
%\texttt{astropy} \citep{astropy:2013,astropy:2018,astropy:2022},
\texttt{DOLPHOT} \citep{dolphin2000b, dolphin2016}

\bibliography{bibliography}{}
\bibliographystyle{aasjournal}

\end{document}